\documentclass{article}

\usepackage{arxiv}

\usepackage[utf8]{inputenc} 
\usepackage[T1]{fontenc}    
\usepackage{hyperref}       
\usepackage{url}            
\usepackage{booktabs}       
\usepackage{amsfonts}       
\usepackage{nicefrac}       
\usepackage{microtype}      
\usepackage{lipsum}
\usepackage{graphicx}
\graphicspath{ {./images/} }

\usepackage{cite}
\usepackage{booktabs} 
\usepackage{xcolor}
\usepackage{amsmath}
\usepackage{amsfonts}
\usepackage{url}

\usepackage{subcaption}

\title{The Greatest Teacher, Failure is: Using Reinforcement Learning for SFC Placement Based on Availability and Energy Consumption}

\author{
 Guto Leoni Santos \\
  Centro de Informática\\
  Universidade Federal de Pernambuco (UFPE)\\
  Recife, Brazil \\
  \texttt{gls4@cin.ufpe.br} \\
  \And
  Theo Lynn \\
  Business School\\
  Dublin City University (DCU)\\
  Dublin, Ireland \\
  \texttt{theo.lynn@dcu.ie} \\
  \And
  Judith Kelner \\
  Centro de Informática\\
  Universidade Federal de Pernambuco (UFPE)\\
  Recife, Brazil \\
  \texttt{jk@gprt.ufpe.br} \\ 
  \And
  Patricia Takako Endo \\
  Universidade de Pernambuco (UPE)\\
  Caruaru, Brazil \\
  \texttt{patricia.endo@upe.br} \\ 
}

\begin{document}
\maketitle
\begin{abstract}
Software defined networking (SDN) and network functions virtualisation (NFV) are making networks programmable and consequently much more flexible and agile. To meet service level agreements, achieve greater utilisation of legacy networks, faster service deployment, and reduce expenditure, telecommunications operators are deploying increasingly complex service function chains (SFCs). Notwithstanding the benefits of SFCs, increasing heterogeneity and dynamism from the cloud to the edge introduces significant SFC placement challenges, not least adding or removing network functions while maintaining availability, quality of service, and minimising cost. In this paper, an availability- and energy-aware solution based on reinforcement learning (RL) is proposed for dynamic SFC placement. Two policy-aware RL algorithms, Advantage Actor-Critic (A2C) and Proximal Policy Optimisation (PPO2), are compared using simulations of a ground truth network topology based on the Rede Nacional de Ensino e Pesquisa (RNP) Network, Brazil's National Teaching and Research Network backbone. 
The simulation results showed that PPO2 generally outperformed A2C and a greedy approach both in terms of acceptance rate and energy consumption. A2C outperformed PPO2 only in the scenario where network servers had a greater number of computing resources.
\end{abstract}


\section{Introduction}
\label{sec:introduction}
%
%
%
%


Historically, telecommunications operators (TOs) deployed special purpose, network-specific, fixed-function hardware with standardized protocols. This hardware-centric approach often relied on fragmented, non-commodity hardware with physical installs for each appliance. As a result, innovation and competition was constrained by both the development of, access to, and maintenance of hardware development \cite{metzler2015}.  The rapid emergence and adoption of the so-called \emph{third IT platform} \cite{gens20133rd}, a convergence of mobile technology, cloud computing, big data analytics and social media, by both enterprises and consumers has resulted in dramatic shifts in bandwidth demand, network infrastructure requirements, and associated economic models that the traditional approaches to networking cannot scale or adapt to easily.  Against the backdrop of increasing competition, TOs need new ways to achieve greater legacy infrastructure utilisation, faster service deployment and new service time-to-market, improved greater agility and flexibility, new revenue streams, while at the same time reducing both operational expenditure (OPEX) and capital expenditure (CAPEX) \cite{lynn2018}.

Software defined networking (SDN) and network function virtualisation (NFV) abstract the implementation of new network functions and decouple them from the hardware infrastructure and associated topological constraints \cite{farshin2019modified}. Consequently, networks become programmable and, as a result, much more flexible and agile. ETSI define NFV as the \emph{“...implementation of network functions in software that can run on a range of industry standard server hardware, and that can be moved to, or instantiated in, various locations in the network as required, without the need for installation of new equipment.”}. By virtualising network functions while retaining the same capabilities as the corresponding physical instances, multiple virtualised network functions can share physical hardware in the form of virtual machines (VMs), a concept that can be massively scaled to encompass large volumes of physical hardware \cite{isg2013network}. As a result, TOs need no longer limit themselves to special-purpose hardware and can achieve the same, if not more, capacity with commercial off the shelf (COTS) commoditised equipment thus reducing both CAPEX and OPEX. Consequently, TOs can benefit from reduced total cost of ownership and operational complexity, real-time bandwidth and network scalability, reduced service deployment and time to market, increased security, and improved visibility and integration \cite{lynn2018}.
Unsurprisingly, there are a myriad of virtual network functions (VNFs) commonly used by TOs. These include firewalls, load balancing, intrusion detection systems (IDS), network address translation (NAT), amongst others \cite{bhamare2016survey,kaur2020comprehensive}.

Service function chaining is defined as an ordered sequence of VNFs and subsequent steering of flows through them to provide end-to-end services\cite{kaur2020comprehensive}. SFCs typically comprise four stages\cite{mirjalily2018optimal}: 

\begin{enumerate}
\item Description – details the functional and non-functional properties of the network services including interfaces and constraints. 
\item Composition – defines the order of network services to compose a functional service. Although these services can work independently, some network functions require a sequenced order to work properly. 
\item Placement - determines in which physical nodes of the infrastructure the virtual functions will be deployed. This stage is essential in the pipeline because the physical resources must be managed optimally in terms of usage to avoid overloading some servers or wasting resources. 
\item Scheduling - defines the time an SFC will be deployed in the infrastructure and the time it will be removed, releasing resources to other deployed SFCs. The scheduling stage is important to minimise the whole execution time of the network services.
\end{enumerate}

Service reliability and availability have been the hallmark for trustworthy computing since the turn of the century \cite{mundie2002trustworthy}. They are the cornerstone of modern SLAs and failure to meet quality of service (QoS) can result in significant financial penalties to TOs as well as lost customers and reputational damage. In the same way that TOs must meet the service level agreements (SLAs) with their customers, SFCs must meet the policy constraints under which the SLA demands including network capacity and acceptable total latency for end users \cite{bhamare2016survey}. While SFCs have many advantages, they are not without challenges.  These challenges are exacerbated by the increasing dynamism and heterogeneity of modern network environments \cite{bhamare2016survey}. In such contexts, the dynamic addition or removal of network functions must be executed in such a way that it does not interfere with the availability and quality of network services while at the same time minimising CAPEX and OPEX. Thus, there is a need for robust, well-tested, dynamic and automatic SFC models, and associated research \cite{bhamare2016survey}.

In this paper, we focus on reinforcement learning (RL) models for dynamic availability- and energy-aware SFC placement. While extant literature has proposed solutions for availability-aware SFC placement, energy-aware SFC placement, and the use of RL for SFC placement, few explore all three in one solution and typically assume static SFC models. We compare two policy-based RL algorithms, Advantage Actor-Critic (A2C) and Proximal Policy Optimization (PPO2), using a simulator built in Python. 
We compare the performance of the algorithms with a greedy approach using a real world network topology based on the Rede Nacional de Ensino e Pesquisa (RNP) network, Brazil’s National Teaching and Research Network backbone.

Our main contributions are:
\begin{itemize}
\item We present a system model for dynamic SFC placement. We model the servers in terms of computing resources, availability levels, and energy consumption. We also define  SFC requests in terms of VNFs (their resource requirements) and the availability requirements determined in SLAs with customers.
\item We propose two RL algorithms for SFC placement. We model the SFC request as an Markov decision process (MDP), where the VNFs from an SFC are allocated in a sequential manner. We also define the environment state, the action representations, and the reward functions which enable the RL agent make the SFC placement. We apply PPO2 and A2C algorithms to solve the SFC placement problem.
\item Our simulation validates that the RL algorithms can be used for SFC placement, achieving good acceptance levels and reducing SFC energy consumption, particularly when compared to a greedy approach.
\end{itemize}

The remainder of this paper is organised as follows. Section \ref{sec:background} provides a brief overview of RL techniques and the conceptualisations of both availability and energy consumption used in the study. In Sections \ref{sec:system-model} and \ref{sec:RL-SFC-placement} we formulate the SFC placement problem as an RBD model and define the representations for the state, action, and reward functions for the SFC placement problem. Section \ref{sec:results} presents the simulation and parameters used to test the proposed algorithms and presents the results. In Section \ref{sec:related-works} we discuss the related works on availability- and energy-aware SFC placement as well as those using RL. The paper concludes with a summary of the paper and future avenues for research in Section \ref{sec:conclusion}.

\section{Background}
\label{sec:background}

\subsection{Measuring Availability}
Dependability relates to the ability of systems to avoid failures and outages of a duration that does not adversely impact the users experience \cite{matos2017redundant}. It is a critical aspect in modern systems, both technically and commercially, from the design of systems (both hardware and software) to their implementation and operation \cite{andrade2017availability}. It is often used as an umbrella term for reliability, availability, security, confidentiality, integrity, and maintainability \cite{costa2016availability}. Although these terms are closely related, they differ from each other. For example, reliability refers to the probability that a system will deliver a service properly until a certain time without any failure \cite{damaso2017integrated}. On the other hand, availability is defined as the percentage of time that the service has functioned properly during the total time of operation, and as such assumes that the system can be repaired during its operation  \cite{araujo2018decision}. In this paper, we will focus on the availability evaluation.

The availability of a system component can be calculated using Equation \ref{eq:av-equation} \cite{fan2017availability}.

\begin{equation}
    \label{eq:av-equation}
    A=\frac{Uptime}{TotalTime}=\frac{Uptime}{Uptime+Downtime}
\end{equation}

Availability can also be defined by the system uptime i.e. how much time the system remains operational divided by the total operation time (the sum of the uptime and the downtime of system). The system uptime can also be defined as the mean time to failure (MTTF), the time when the system is not at fault and working properly. Conversely, system downtime can be defined as the mean time to repair (MTTR), that is, the mean time the system fails and is under repair. Thus, availability can be defined as: 

\begin{equation}
    A=\frac{MTTF}{MTTF+MTTR}
\end{equation}

\begin{figure}[!htb]
 \centering
  \includegraphics[width=0.9\columnwidth]{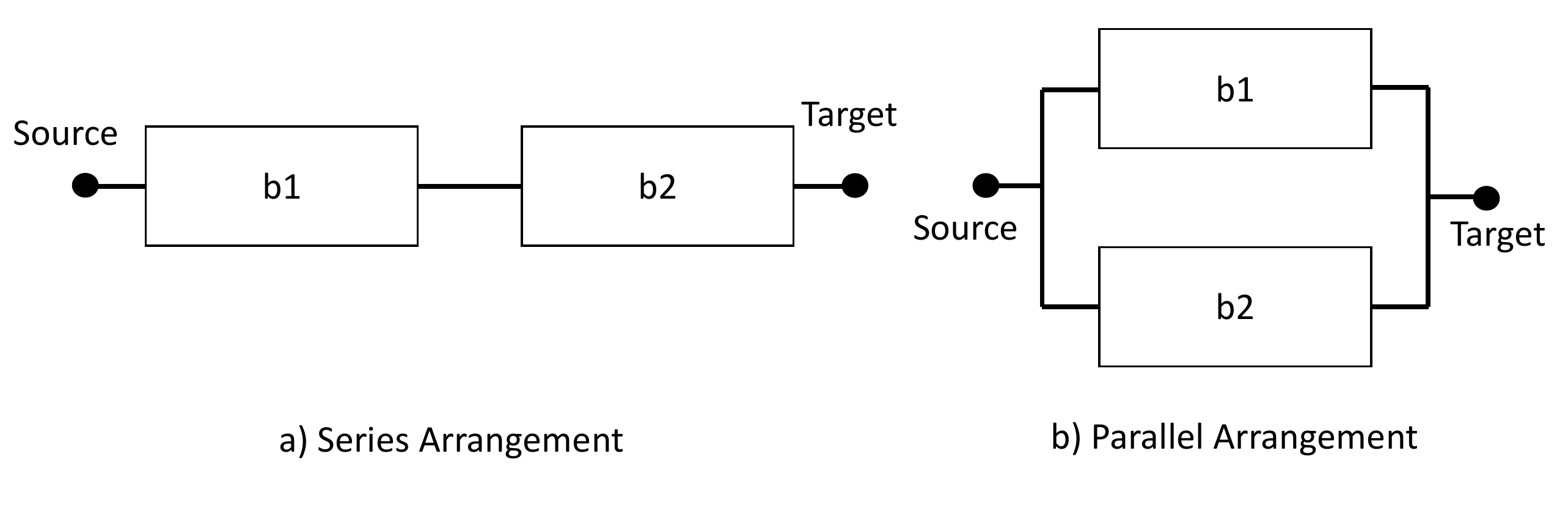}
  \caption{RBD configurations: (a) series and (b) parallel}
  \label{figure:rbdmodel}
\end{figure}

Numerous techniques exist for evaluating the availability of complex systems, such as stochastic Petri nets, continuous Markov chains, and fault trees \cite{damaso2014reliability}. In this paper, we make use of the reliability block diagrams (RBD). RBD is a mathematical formalism to define the relationships of system components. It is frequently used to assess the availability, maintainability, and reliability of these systems \cite{gissler2015system}. System components are represented as a graphical structure composed of connected blocks. In essence, these components can be arranged in two different ways: series and parallel \cite{santos2018analyzing}, as shown in Figure \ref{figure:rbdmodel}.

In a system with a series configuration, all components need to be working for the whole system to remain operational. In a system with a parallel configuration, if one component is working, the whole system remains operational. Equation \ref{eq:series} defines the availability of a series system, $A_s$, where $A_i$ is the availability of each component, that can be calculated through Equation \ref{eq:av-equation} \cite{hoyland2009system}.

\begin{equation}
\label{eq:series}
    A_s = \prod_{i=0}^{N} A_i,
\end{equation}

The availability of a parallel system, $A_p$, can be defined by Equation \ref{eq:parallel}, where the availability of the components is $A_i$ \cite{hoyland2009system}.

\begin{equation}
\label{eq:parallel}
    A_p = 1 - \prod_{i=0}^{N} (1 - A_i),
\end{equation}

RBD can be used to model different SFC configurations and assess their availability. Usually, SFCs are deployed as sequential functions \cite{cai2020composing}, and can be considered a series system, where the failure of any function makes the whole SFC unavailable. Adding redundant functions (in a parallel configuration) can increases the availability of an SFC, however, these additional components also increase the system costs, a key consideration in SFC placement. RL is an option for planning SFC placement that takes into account the actual network resources as well as others metrics of interest \cite{li2019efficient}, including availability.

\subsection{Reinforcement Learning}
\label{subsec:reinforcement-learning-bkg}
RL is a learning paradigm that has featured more prominently in recent literature. It refers to the use of autonomous software agents that learn to perform a task by trial and error, without any human intervention \cite{goodfellow2016deep}. In other words, in RL, the learning process happens through the iterative interaction of the agent with the environment. As such, unlike Deep Learning, RL does not use a fixed data set for training, validation and testing  \cite{troia2019reinforcement}. Instead, RL uses the feedback from the environment in which the agent is inserted. In the learning process, the agent tries to maximise a reward by observing the consequences of its actions \cite{arulkumaran2017brief}. 

The tasks performed by the RL agent can be classified as episodic tasks or continuing tasks \cite{ravichandiran2018hands}. In episodic tasks, there are one or more final states. When the agent achieves a given state, the task restarts. To use an analogy, the agent is like a driver in a racing car video game where each race is an episode. The agent starts each race and continues until the end of a race, concluding an episode. When the race is over, the agent can start again from the initial state; each episode is independent of the others. In continuing tasks, there is no terminal state. The agent actuates in the environment achieving rewards indefinitely, more like a robot that continuously responds to commands and completes tasks. 

RL is supported by a formalism called MDP. MDP is composed of \cite{arulkumaran2017brief}:

\begin{itemize}
    \item A finite set of states $\mathcal{S}$, plus a distribution of starting states $p(s_0)$.
    \item A finite set of actions $\mathcal{A}$.
    \item A set of dynamic transition $\mathcal{T}(s_{t+1}|s_t, a_t)$ that maps a state to an action at time \textit{t} onto a distribution of states at time $t+1$.
    \item A reward function $\mathcal{R}(s_t,a_t,s_{t+1})$.
    \item A discount factor $\gamma \in [0,1]$, where lower values means more immediate rewards.
\end{itemize}

In general, a policy $\pi(a_t|s_t)$ defines the agent behavior. At instant \textit{t}, the policy maps an action $a_t$ into a state $s_t$. A reward $r_t$ is calculated, and transitions to the next state $s_{t+1}$ following a state transition probability $\mathcal{P}(s_{t+1}|s_t,a_t)$. Considering an episodic MDP, this process continues until the terminal state is reached. For each step, the reward is accumulated from the environment resulting into the returned value $R=\sum_{t=0}^{T-1}\gamma^{t}r_{t+1}$. The main goal of RL is to find the optimal policy $\pi^{*}$ (Equation \ref{eq:expected-return}) that results in the maximum expected return (reward) for all states.

\begin{equation}
\label{eq:expected-return}
\underset{\pi}{\pi^{*}=\mathtt{argmax} \mathbb{E}[R|\pi]}
\end{equation}


A widely-used class of algorithms in the literature are value-based methods \cite{osband2016deep,mousavi2017traffic,pan2018multisource}.
These algorithms try to extract the near optimal policy based on the value function, which is defined in Equation \ref{eq:value-equation}. The value ($V$) is the expected long-term reward achieved by the policy ($\pi$) from a state $s$.

\begin{equation}
\label{eq:value-equation}
    V^{\pi}(s_t)=\sum_{i=0}^{\infty}\gamma^iR_{t+i}
\end{equation}

The Q-learning is a family of algorithms which learn how to optimize the quality of an action ($Q$ value).
Equation \ref{eq:q-simple} defines the $Q$ value of an action $a$ in a given state $s$ following a policy $\pi$ at time $t$.

\begin{equation}
\label{eq:q-simple}
    Q^{\pi}(s,a)=\mathbb{E}_\pi[R_t|s_t=s,a_t=a]
\end{equation}

The best policy is derived from the $Q$ value of each state, e.g. select the action which returns the maximum expected reward for a given state. The $Q$ value is learned in an interactive way during agent training. The Bellman Equation (Equation \ref{eq:q-update}) updates the $Q$ value during the training \cite{hasselt2010double}.

\begin{equation}
    \label{eq:q-update}
    Q(s_t,a_t)\leftarrow Q(s_t,a_t)+\alpha(r_{t+1}+\gamma\underset{a}{max}Q_t(s_{t+1},a)-Q_t(s_t,a_t))
\end{equation}

The new estimation of $Q$ value is the sum of the old estimation and the error. The error is defined by the reward achieved, $r_{t+1}$, plus the difference between the new $Q$ value obtained, $\underset{a}{max}Q_t(s_{t+1},a)$ minus the old value, $Q_t(s_t,a_t)$. $\gamma$ is the discount factor and is a value between 0 and 1 t that defines the importance of the immediate reward. $\alpha$ is the learning rate that defines how much the $Q$ values updated.









\begin{figure}[h]
\centering
\includegraphics[width=0.5\columnwidth]{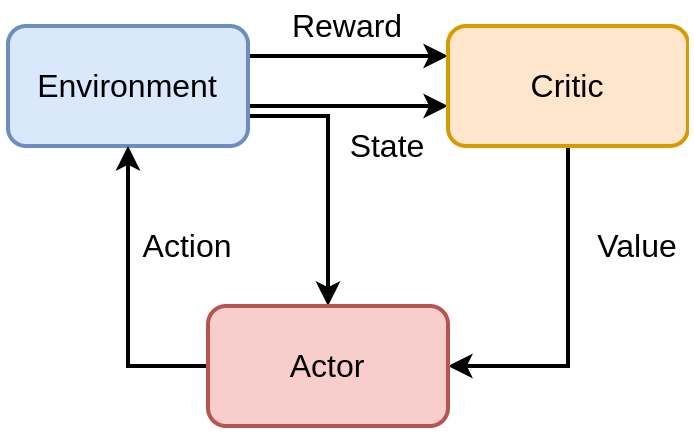}
\caption{Actor-critic method}
\label{fig:actor-critc-method}
\end{figure}

Another method to find the policy $\pi$ is called \textbf{Actor Critic} \cite{sutton2018reinforcement}. Figure \ref{fig:actor-critc-method} shows the architecture of a traditional Actor Critic algorithm. The actor is responsible to choose the action which guides the agent; the critic learns the quality of the actions from actor the based on the reward from the environment i.e. the $Q$ value.

The actor and critic are two independent functions as shown in Equation \ref{eq:policy-function} and Equation \ref{eq:value-function}, respectively. These functions have parameters, $\theta$ for actor function and $w$ for the critic function, which are learned during the training.

\begin{equation}
    \label{eq:policy-function}
    \pi(s,a,\theta)
\end{equation}

\begin{equation}
    \label{eq:value-function}
    q(s,a,w)
\end{equation}

For each time step $t$, the state is forwarded for the actor and critic. The actual policy takes the state and outputs an action $a$, which results in a new stat $S_{t+1}$ and a reward $r_{t+1}$. Afterwards, the critic calculates the $Q$ value ($q_w$) and the actor updates its parameters using the $Q$ value using Equation \ref{eq:policy-update}. After updating its parameters, the actor produces a new action and the critic updates its parameters using Equation \ref{eq:value-update}. The actor and critic functions have different learning rates, $\alpha$ and $\beta$, respectively.

\begin{equation}
\label{eq:policy-update}
\Delta\theta=\alpha\nabla_\theta(log\pi_\theta(s,a))q_w(s,a)
\end{equation}

\begin{equation}
    \label{eq:value-update}
    \Delta w=\beta(r(s,a)+\gamma q_w(s_{t+1},a_{t+1})-q_w(s_t,a_t))\nabla_w q_w(s_t,a_t)
\end{equation}



\subsubsection{Advantage Actor-Critic (A2C)}

The A2C method is variant of the actor-critic method which includes an advantage concept derived from the Q value as shown in Equation \ref{eq:advantage-definition}. While the state value $V$ defines how good is to be at a state (Equation \ref{eq:value-equation}), in effect the expected return from a policy \cite{sutton2018reinforcement}, the advantage $A$ defines how good an action is when compared with other actions. In A2C, the RL agent uses the advantage value instead of the Q value during training, which reduces the high variance of the policy networks and stabilizes the model training \cite{peng2018adversarial}.

\begin{equation}
\label{eq:advantage-definition}
\begin{split}
    Q(s,a)=V(s)+A(s,a)\\
    A(s,a)=Q(s,a)-V(s)\\
    A(s,a)=r+\gamma V(s_{t+1})-V(s)
\end{split}
\end{equation}

In the A2C algorithm, multiple agents explore different parts of the environment in parallel thus increasing exploration efficiency compared to a single agent \cite{mnih2016asynchronous}. The parameters of all the agents are used to update a set of global parameters. A coordinator controls this update, which happens after each agent finishes its exploration. Then, in the next iteration, all agents use the same global parameters to explore the environment. This parameter synchronization makes the training more cohesive and potentially faster.


\subsubsection{Proximal Policy Optimization (PPO)}

PPO combines the idea of having many agents from A2C with the idea of exploring the policy region in the optimization problem during the training \cite{schulman2017proximal}. Instead of using the log of $\pi$ to update the policy parameters (Equation \ref{eq:policy-update}), PPO uses the ratio between the probability of action under the current policy ($\pi_\theta$) divided by the probability of action under the previous policy ($\pi_{\theta old}$). Equation \ref{eq:ratio-policy-equations-ppo} describes the ratio.

\begin{equation}
    \label{eq:ratio-policy-equations-ppo}
    r_t(\theta)=\frac{\pi_\theta(a_t|s_t)}{\pi_{\theta old}(a_t|s_t)}
\end{equation}

The policy loss in PPO can be defined as Equation \ref{eq:ppo-loss-function}. 

\begin{equation}
    \label{eq:ppo-loss-function}
    L(\theta)=\mathbb{\hat{E}}_t=\begin{bmatrix}
                    \frac{\pi_\theta(a_t|s_t)}{\pi_{\theta old}(a_t|s_t)}\hat{A}_t
                    \end{bmatrix}
\end{equation}

If the variation probability from the previous policy compared to the current policy is too high, it can cause an excessive policy update resulting in a large policy gradient step.In order to mitigate the exploding gradient problem resulting from  a large step, PPO uses the clipped surrogate objective function (Equation \ref{eq:clip-function-ppo}).

\begin{equation}
    \label{eq:clip-function-ppo}
    L^{CLIP}(\theta)=\mathbb{\hat{E}}[min(r_t(\theta)\hat{A}_t,clip(r_t(\theta),1-\epsilon,1+\epsilon)\hat{A}_t)]
\end{equation}

As per Equation \ref{eq:clip-function-ppo}, we have two probability ratios, one non clipped and one clipped between $1-\epsilon$ and $1+\epsilon$, where $\epsilon$ is a hyper parameter which defines the clip range. The minimum between these ratios is taken as the final objective in the PPO algorithm, which is the lower (pessimistic) bound of the unclipped objective.

\section{System Model}
\label{sec:system-model}
We consider an infrastructure composed of a set of commodity servers, $S$, to host the VNFs of a SFC request. We assume that a server has connectivity to any other server in the network, even not directly.

\begin{figure}[h]
\centering
\includegraphics[width=0.8\columnwidth]{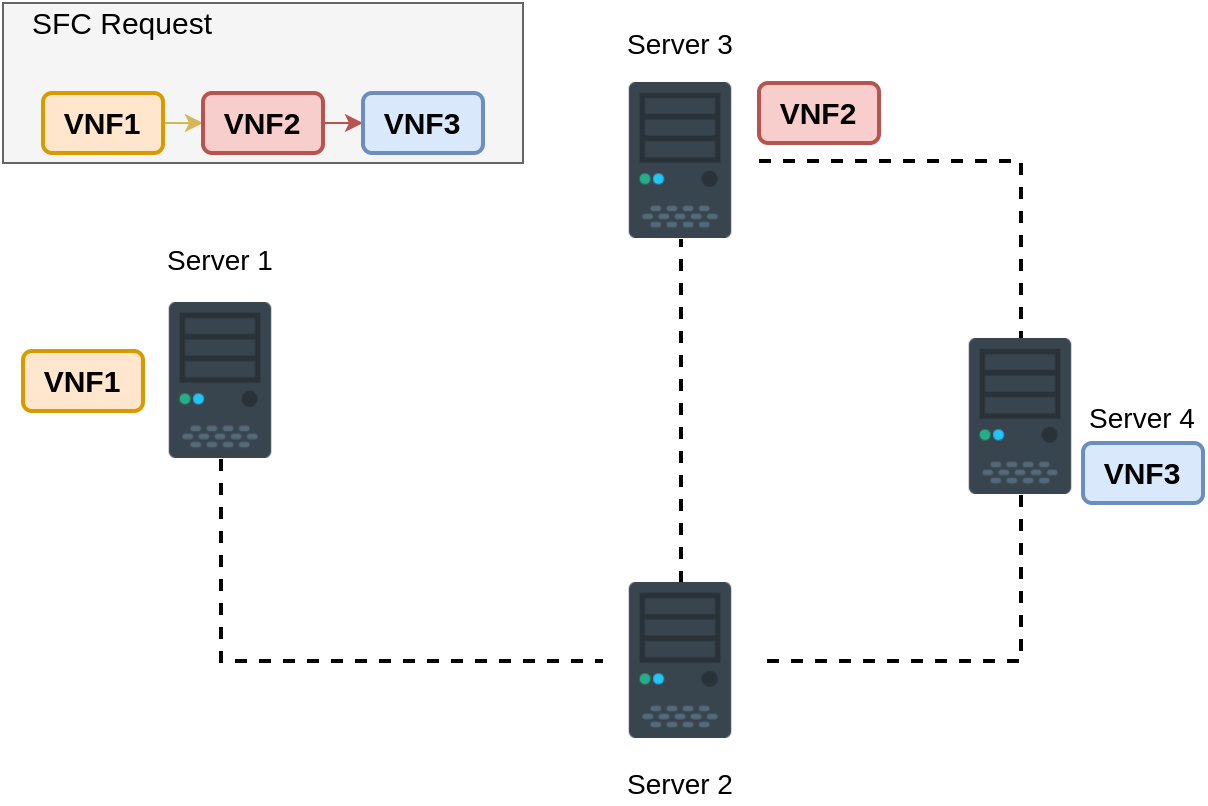}
\caption{Simple network example}
\label{fig:network-example}
\end{figure}

A simple scenario with four servers ($|S|=4$) is illustrated in Figure \ref{fig:network-example}. The Server 1 is connected to the Server 4 through the Server 2 or through the Server 2 and the Server 3. Therefore, the VFN1 placed on Server 1 has connectivity to the VNF2 and VFN3, hosted on the Servers 3 and 4, respectively. 

Each server may fail during its operation time. We consider that hardware failures can have different causes, such as memory failure, CPU overheating, storage failure, and so on. These failures can happen following a stochastic process, which is described by a probability distribution function (PDF). The parameters of the PDF describe the behaviour of failures and can be adjustable to represent the different types of failures. These servers may also be repaired after a failure. The total repair time includes the failure detection time and the repair operation time, but in this paper we consider as an unique time. The repair time also follows a stochastic process and is defined by a PDF with the respective parameters.

These servers have a limited capacity to deploy VNFs, which is related to the computational resources such as CPU and memory. The resources of the $i^{th}$ server is represented by $r_i$. Each server has an energy consumption associated regarding its operation. We assume that a customer can have a list of SFC templates. An SFC template defines the VFNs and the respective order given the client requirements. A VFN is defined by its functionalities and resource requirements. An SFC definition can have different VFNs with different resource requirements, which increases the complexity of the placement process. Similar to the servers, we assume that the  VFNs are deployed as virtual machines or containers, and can fail and be repaired during their operation time. The VFN failure and repair events are also represented by PDFs with their respective parameters. 

Due to server and VFN failures, SFC availability is impacted. As each customer defines an availability level which must be met to avoid breaking the SLA, SFC unavailability can result in SLA violations and potential financial penalties or reputational damage for the service provider. In this paper, we use RBD to model the SFC components (servers and VFNs) and calculate the overall SFC availability.

\begin{figure}[ht]
\centering
\includegraphics[width=0.8\columnwidth]{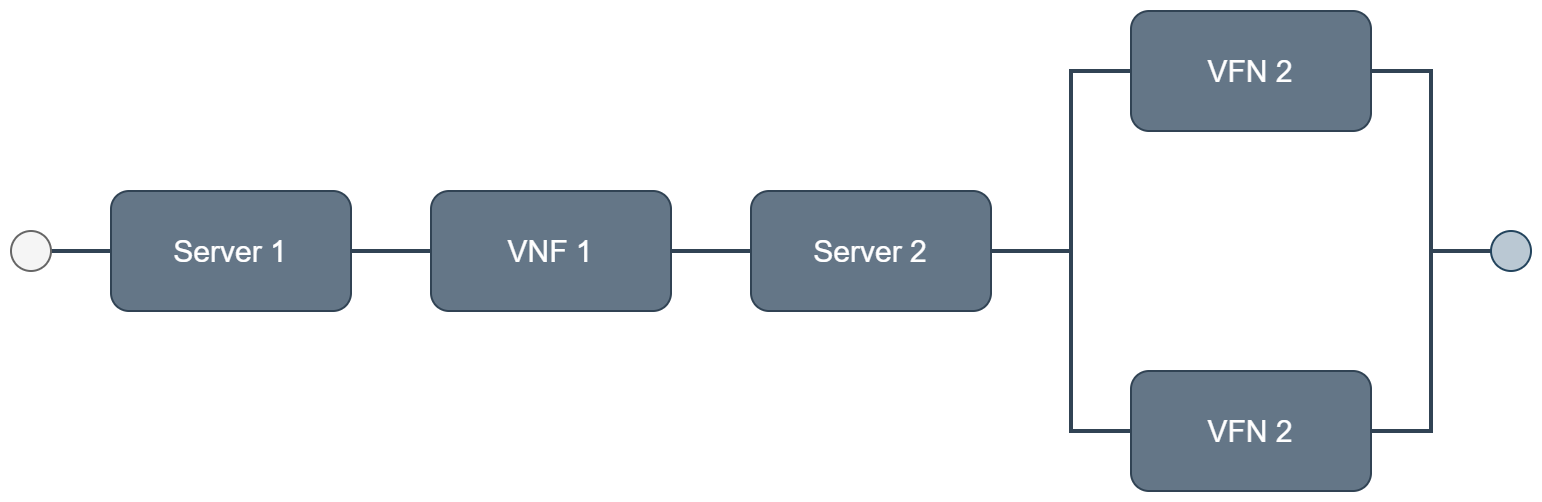}
\caption{An example of RBD of an SFC.}
\label{fig:rbd-example}
\end{figure}

Figure \ref{fig:rbd-example} shows an example of a RBD that represents an SFC composed of two VFNs (VNF1 and VNF2) deployed in two servers (Server 1 and Server 2). One instance of VFN1 is deployed in Server 1 while two redundant instances of VFN2 are deployed in Server 2. In this example, the whole SFC is considered operational when Server 1 and the VNF1 instance are operational, as well as Server 2 and at least one instance of VNF2. Using this notation, different SFC placements can be represented and its availability can be evaluated.

\section{RL for SFC placement}
\label{sec:RL-SFC-placement}

In this section, we describe how we use RL algorithms for SFC placement. Firstly, we describe the characteristics of SFC requests and the RL formulation for the SFC placement problem. We define the environment state representation, the action representation, and the reward function.

\subsection{Characteristics of SFC Requests}
\begin{figure}[h]
\centering
\includegraphics[width=1\columnwidth]{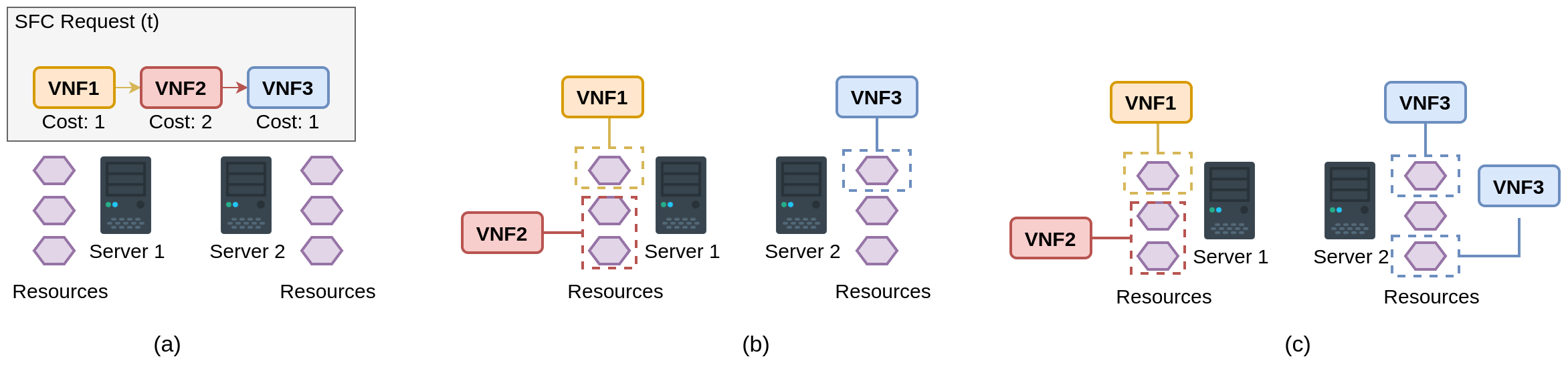}
\caption{Example of an SFC request.}
\label{fig:sfc-request-example}
\end{figure}

Our SFC requests formulation is based on \cite{xiao2019nfvdeep} and is illustrated in Figure \ref{fig:sfc-request-example}. An SFC request arrives at time \textit{t}, and is composed of three VFNs of different types (VFN1, VFN2, and VFN3), having resource costs equal to 1, 2, and 1, respectively. For simplicity, the SFC availability requirements are set at the same level the service level agreement (SLA) agreed with the customer; in this case we set it at 95\% (hereinafter referred to as the availability requirement). The infrastructure is composed of two servers, Server 1 and Server 2, each with three computational resources.

We formulate our SFC placement problem as an MDP, where the states represents the resources available in all servers present in the system at time \textit{t}. Initially, Server 1 and Server 2 each three available resources. The request is divided into sub-requests according to the number of VFNs that compose the SFC; in this example, there are three sub-requests. Then, each sub-request is processed, allocating the VNFs to be deployed in one or more servers. Multiple VNF instances of the same type may be deployed in more than one server in order to increase the whole SFC availability.

Each VNF is processed in a sequential manner, until the whole SFC is placed. Figure \ref{fig:sfc-request-example}b illustrates an SFC placement. VNF1 and VFN2 are deployed on Server 1, consuming all resources available in this server. VFN3 is deployed on Server 2, after which only two resources remain. 

Figure \ref{fig:sfc-request-example}c illustrates an SFC placement with two VFN3 instances deployed on Server 3. This strategy may be adopted if high availability is required under the SLA. In this scenario, redundant VNF instances may be deployed to ensure the SLA is met. This can exacerbate placement complexity as the placement depends on a number of factors including the network topology, the remaining resources, different VFN types, and the request distribution \cite{zhang2018enabling}.

Selecting the best server in which to deploy a VFN is not a trivial task, even if just considering availability and the allocation of redundant instances. Increasing availability through redundancy may increase overall system energy consumption and, as a result, OPEX. Indeed, a number of studies suggest that the SFC placement problem is NP-hard \cite{cai2020composing, zhang2019near} requiring heuristic or meta-heuristic solutions \cite{tavakoli2019multi}. The possibility of server failure increases the complexity of the placement problem as server failure reduces the total system capacity to allocate SFCs. As such, we propose RL algorithms to select the best servers and the redundancy schema for VFN instances. The RL agents learn by interacting with the environment, taking actions and receiving rewards \cite{lesort2018state}. Based on the status of the servers, in terms of available resources and SFC requirements, the agent can learn the placement policy that maximises the number of SFC requests considering the required availability as per the SLA. 

The following subsections describe the representations for the state, action, and reward functions for the SFC placement problem.

\subsection{Environment State Representation}

The state representation is the set of information that the RL agent receives from the environment to take an action that will return a given reward in the long run \cite{sutton2018reinforcement}. This information must be relevant for the agent to take the right action for a given state. The state representation at time \textit{t} ($S_t$) is composed of four components as defined by Equation \ref{eq:state-representation}.

\begin{equation}
    \label{eq:state-representation}
    S_t=\{R_i,A_i,\omega_c,\theta_{c,j}\}, 1\leq i \leq|S|, c \in 1,2,3,...,C
\end{equation}

where $R_i$ represents the remaining resources of the server \textit{i}; $A_i$ is the availability level of the server $i$; $\omega$ is the resources required by the current VNF, \textit{j}, required by the customer, $c$; and $\theta_c$ is the SFC availability requirement by the customer, $c$. 

\subsection{Action Representation}

The action representation is a vector that defines a unique action that the RL agent can take at a given time \textit{t}. In this paper, the action is the SFC placement decision for each VNF. Equation \ref{eq:action-representation} shows the action representation used in this study.

\begin{equation}
    \label{eq:action-representation}
    A_t=\{s_i,Q\} 1\leq i \leq |S|, 1\leq Q \leq VM_{max}
\end{equation}

The action taken at time \textit{t}, $A_t$, is composed of two numbers. The former, $s_i$, is the identifier of a server from the set of servers available in the system; the latter, $Q$, is the quantity of redundant VNF instances that will be created in the server $s_i$. For simplicity, we assume that all redundant VNFs of the same type from an SFC request are allocated to the same server. We also assume that, for each VNF, there is a maximum number of redundant instances, $VM_{max}$, allowed to be allocated. It is worth noting that if more servers are added in the architecture, the action representation is not affected. As such, this notation works for scenarios with hundreds or thousands of servers in contrast to representations where all possible servers combinations are considered as action representations \cite{chai2019parallel}.

\subsection{Reward Function}

The reward is a signal that the RL agent receives from the environment after taking an action in a specific state. Based on the reward value, the agent learns the best policy to follow during training \cite{sutton2018reinforcement}. The reward considers SFC availability (when it is allocated in the infrastructure, considering hardware and software components), the availability requirement for an SFC request, and the total energy consumption of the allocated SFC. The energy consumption of an SFC is calculated as the sum of the energy consumption of all servers where its VNFs are allocated \cite{sayadnavard2019reliable}. The reward is calculated as per Equation \ref{eq:reward}.

\begin{equation}
    \label{eq:reward}
        R(s_t,a) = 
        \begin{cases}
            -1 & \mbox{not available resources} \\
            0 & \mbox{available resources} \\
            ((SFC_{av}-\theta_{c,i})*\varrho-energy_i*\varsigma)+ 2, & \mbox{r is accepted } \\ 
            -5, & \mbox{r is rejected } \\
        \end{cases}
\end{equation}

If the agent is allocating an intermediate VNF (i.e. not the final VNF of an SFC request), the agent does not receive a reward. However, if the agent tries to allocate a VNF in a server with insufficient resources, the agent receives a penalty (a negative reward) of -1. When the SFC request is accepted (i.e. when all VFNs required are allocated), the reward $R$ for a state $s$ at time $t$ after the agent takes the action $a$, is defined by the difference between the availability of the SFC allocated by the RL agent and the availability requirement defined by the customer $c$ for the SFC $i$. This difference is multiplied by a factor $\varrho$. In addition, we consider the energy consumption of the SFC ($energy_i$) multiplied by the factor $\varsigma$. The factors $\varrho$ and $\varsigma$ can be used to adjust the scale of the availability difference or the energy consumption and, if needed, give greater weight to either the availability of the allocated SFC or the SFC energy consumption. If the SFC request is rejected (i.e. if the RL agent tries to allocate a VNF in a server without available resources), the agent again receives a penalty, -5.

Using the reward defined by Equation \ref{eq:reward}, the RL agent is incentivised to allocate SFCs with availability higher than the customer requirements. As noted, the RL agent can allocate SFCs with availability lower than the requirement but it will result in a penalty for the agent.



One option to achieve a higher availability is to allocate many redundant VNF instances but this comes with a cost related to increased energy consumption. Thus, we subtract the energy consumption of the SFC allocated by the agent from the reward. Since the availability and the energy consumption are in different scales, we use the factors $\varrho$ and $\varsigma$ to put them in the same scale. 

Where the SFC is not allocated, the RL agent receives a large penalty, since it is an option that should be avoided as much as possible.

\section{Evaluation}
\label{sec:results}
In this section, we present the results of the SFC placement using two RL algorithms, A2C and PPO. Firstly, we describe the simulation scenario and the parameters used in the experiments. Then, we present the results of the algorithms parametrization. Finally, we present the results of algorithms for different scenario configurations and compare them against a greedy algorithm.

\subsection{Simulation Setup}
\label{subsec:simulation-setup}

To perform the experiments in this paper, we built a simulator using the Python language to generate the SFC requests, server failures, and repair events. The arrival of SFC requests follows a Poisson process with a constant arrival rate $\lambda$. The lifetime of SFC requests follows an exponential function with rate $1/\mu$, where $\mu$ is the average lifetime \cite{li2018context}. To represent a diverse range of customers, each customer has their own arrival rate and average SFC lifetime. 

The failure and repair events of the servers and VFNs follow an exponential function with failure rate and arrival rate, respectively \cite{torquato2019iaas,lima2020data}. However, others PDFs can be used, if required. The failure rate is defined by $1/MTTF$, where $MTTF$ corresponds to either to the servers or the VNFs. The repair rate is defined by $1/MTTR$, where the servers and the VNFs each have their respective values. 
It is important to highlight that our simulator allows these parameters to be easily changed to represent different scenarios.

\begin{table}[ht]
\caption{Simulation Parameters}
\centering
\label{tab:simulation-parameters}
\begin{tabular}{@{}cc@{}}
\toprule
\textbf{Parameter}                  & \textbf{Value} \\ \midrule
Number of servers                   &28               \\
Number of resources per server      &10                \\
Server MTTF                         &8760h  \cite{araujo2014availability}               \\
Server MTTR                         &1.667h \cite{araujo2014availability}               \\
VNF MTTF                            &2880h  \cite{araujo2014availability}              \\
VNF MTTR                            &0.17h \cite{araujo2014availability}               \\
Energy consumption (CPU)            &40W \cite{ali2015energy}               \\
Energy consumption (memory)         &30.17W  \cite{ali2015energy}              \\
Maximum number of redundant VNFs     &4                \\
Number of customers                 &5                \\
Availability Requirement (SLA) & 99.9\%       \\ 
Arrival rate ($\lambda$)            & 0.04 request/h \cite{palhares2014joint}               \\
SFC lifetime ($\mu$)                 &1000h \cite{palhares2014joint}  \\
\bottomrule
\end{tabular}
\end{table}

\begin{figure}[h]
\centering
\includegraphics[width=0.5\columnwidth]{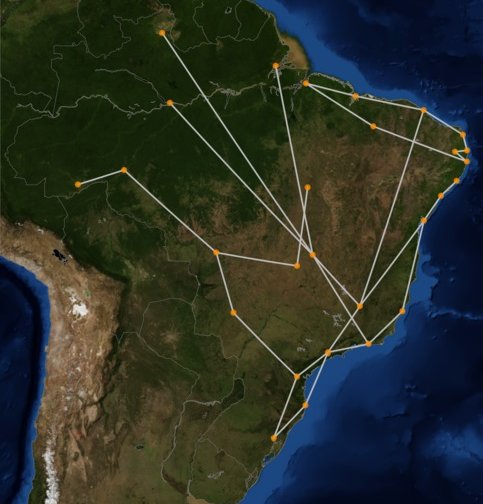}
\caption{RNP network}
\label{fig:rnp}
\end{figure}

Firstly, we consider a basic simulation scenario based on the assumptions summarised in Table \ref{tab:simulation-parameters}. We assume a network with 28 physical servers based on the RNP and illustrated in Figure \ref{fig:rnp}. Each server has 10 computational resources available to deploy VFNs from SFC requests. We assume server MTTF and MTTR of 8760 hours and 1.667 hours, respectively. For the VNFs, we assume parameters of generic virtual machines with an MTTF of 2880 hours and an MTTR of 0.17 hours as per \cite{araujo2014availability}. In line with \cite{ali2015energy}, we assume CPU energy consumption of 40W and memory energy consumption of 30.17W. The energy consumption increases linearly as new servers are added in the SFC placement. The maximum number of redundant VNFs of the same type is four. Consistent with industry practice, we assume that SFC availability requirement is 99.9\%.

The arrival rate $\lambda$ is four SFC requests per 100 hours, while the SFC lifetime is 1000 hours \cite{palhares2014joint}.
In a simulation, an SFC is generated randomly for each customer composed of two or three VNFs from the types previously defined; $\lambda$ and $\mu$ parameters are generated based on the standard values described previously. To generate these parameters, we consider a uniform distribution with lower and upper bounds equal to plus and minus 10\% of the standard value of $\lambda$ and $\mu$.

We assume a fixed number of customers of five. We assume three different VNF types - a Wide Area Network-optimiser (WAN-opt), a firewall, and an Intrusion Detection System (IDS). The firewall and IDS require one computational resource from the server, while WAN-opt requires four computational resources from the servers \cite{xu2018energy}. 


In this study, we compare two different RL algorithms, A2C and PPO. In the following subsections we describe the results of parameterization experiments and the application of these algorithms in different scenarios.

\subsection{Parameterization Results}
First, the algorithms were configured with their respective parameters. We varied two common parameters of both algorithms that impact their learning process - the learning rate ($\alpha$) and the discount factor ($\gamma$). 

The algorithms were implemented using the Stable Baselines library\footnote{\url{https://stable-baselines.readthedocs.io/en/master/index.html}}. Stable Baselines is a Python library that provides a set of improved implementations of RL algorithms based on OpenAI Baselines. The source code used to perform the experiments of this study is available at GitHub\footnote{\url{https://github.com/GutoL/SFC_RL}}. 
The PPO used in this study is the second version named PPO2 provided by Stable Baselines framework.
The other parameters were provided by the Stable Baselines\footnote{\url{https://stable-baselines.readthedocs.io/en/master/modules/a2c.html}}\footnote{\url{https://stable-baselines.readthedocs.io/en/master/modules/ppo2.html}}. The variation of $\alpha$ and $\gamma$ were based on the standard values, and are shown in Table \ref{tab:alpha-gamma-variation}.

\begin{table}[h]
\caption{$\alpha$ and $\gamma$ variation for parametrization}
\label{tab:alpha-gamma-variation}
\centering
\begin{tabular}{@{}cc@{}}
\toprule
Parameter & Variation                             \\ \midrule
$\alpha$  & 0.00005, 0.00025, 0.0005, 0.00075 \\
$\lambda$ & 0.85, 0.9, 0.99, 0.95            \\ \bottomrule
\end{tabular}
\end{table}

\begin{figure*}[h]
\centering
    \begin{subfigure}[b]{0.4\textwidth}
    \includegraphics[width=\textwidth]{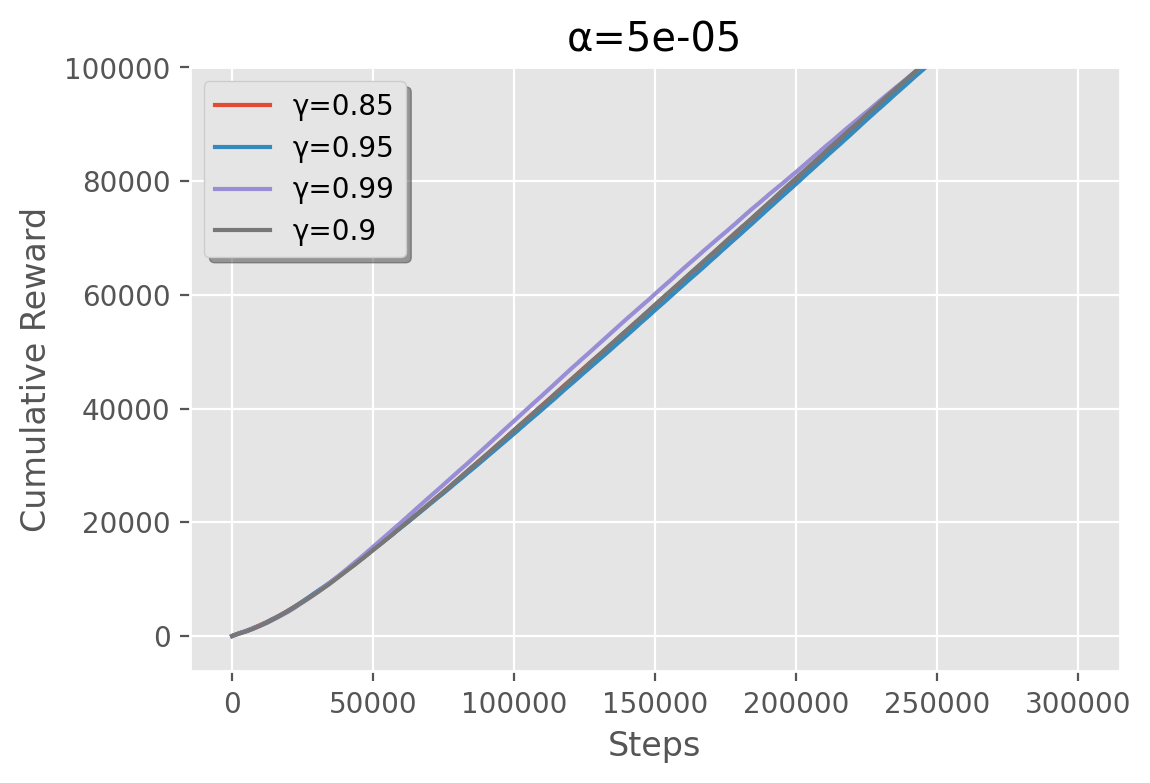}
    \caption{}
    \end{subfigure}
    \begin{subfigure}[b]{0.4\textwidth}
    \includegraphics[width=\textwidth]{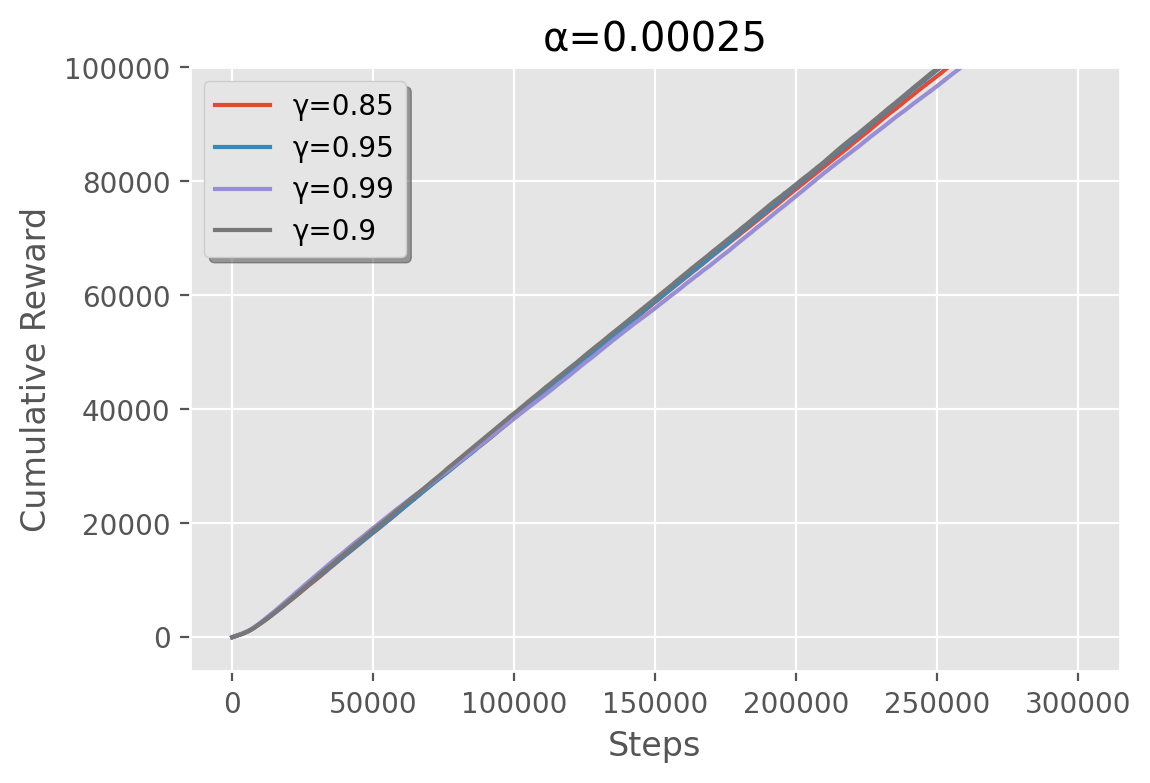}
    \caption{}
    \end{subfigure}

    \begin{subfigure}[b]{0.4\textwidth}
    \includegraphics[width=\textwidth]{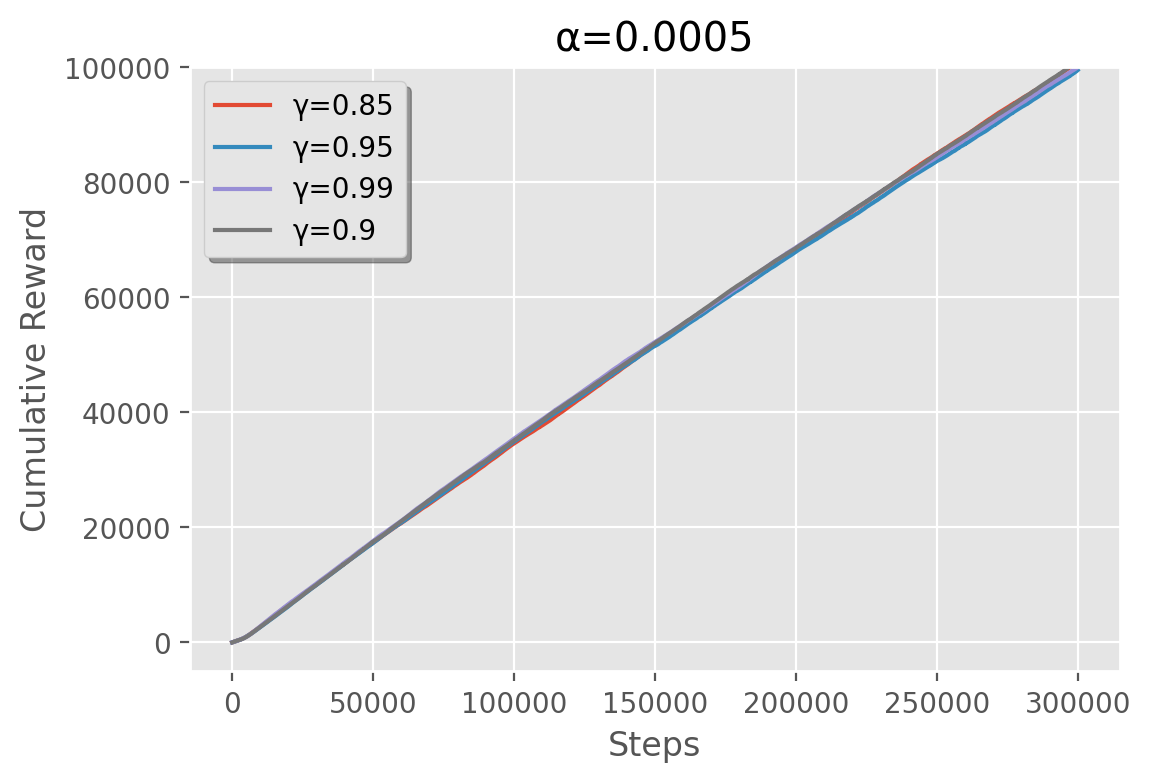}
    \caption{}
    \end{subfigure}
    \begin{subfigure}[b]{0.4\textwidth}
    \includegraphics[width=\textwidth]{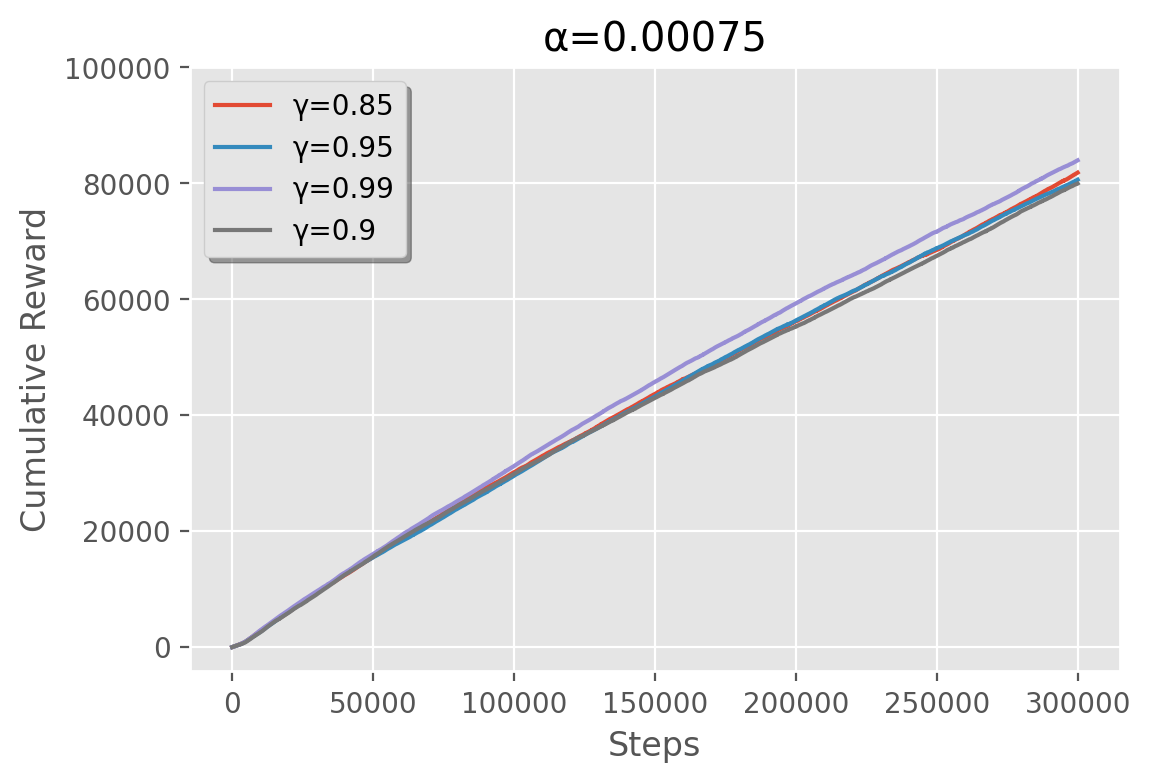}
    \caption{}
    \end{subfigure}
\caption{Parameter variation of A2C algorithm}
\label{fig:parametrization-a2c}
\end{figure*}

\begin{figure*}
\centering
    \begin{subfigure}[b]{0.4\textwidth}
    \includegraphics[width=\textwidth]{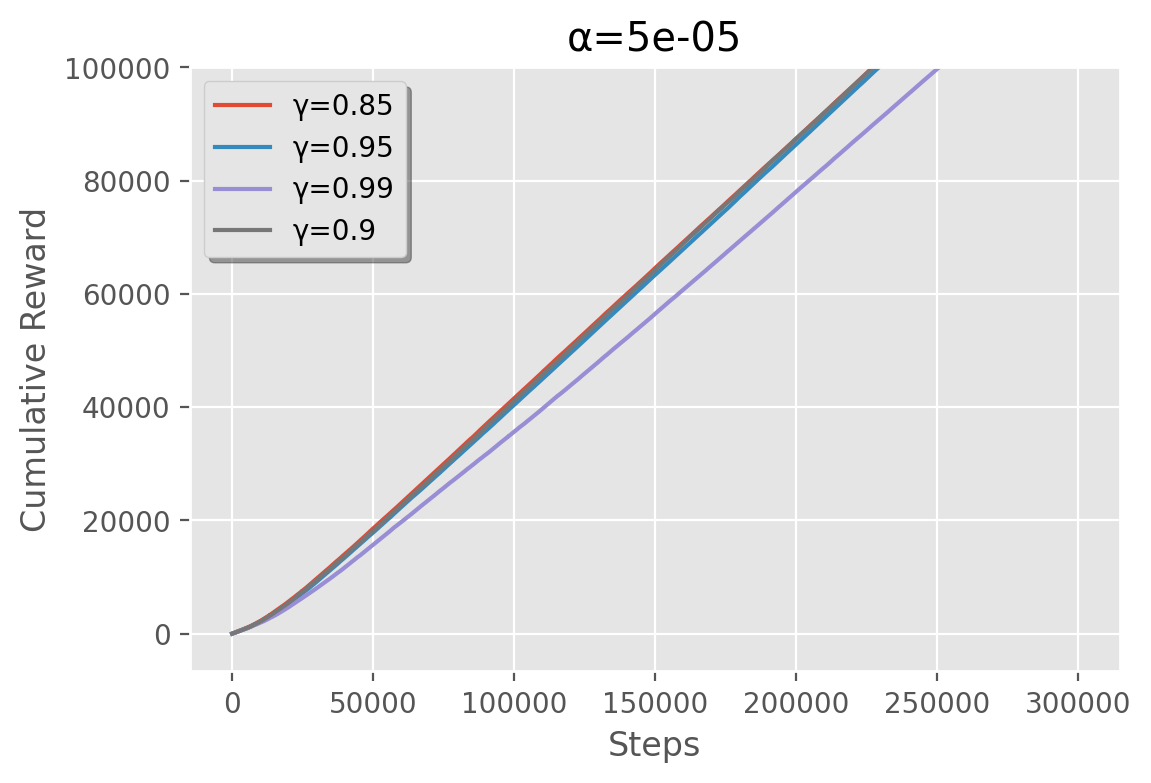}
    \caption{}
    \end{subfigure}
    \begin{subfigure}[b]{0.4\textwidth}
    \includegraphics[width=\textwidth]{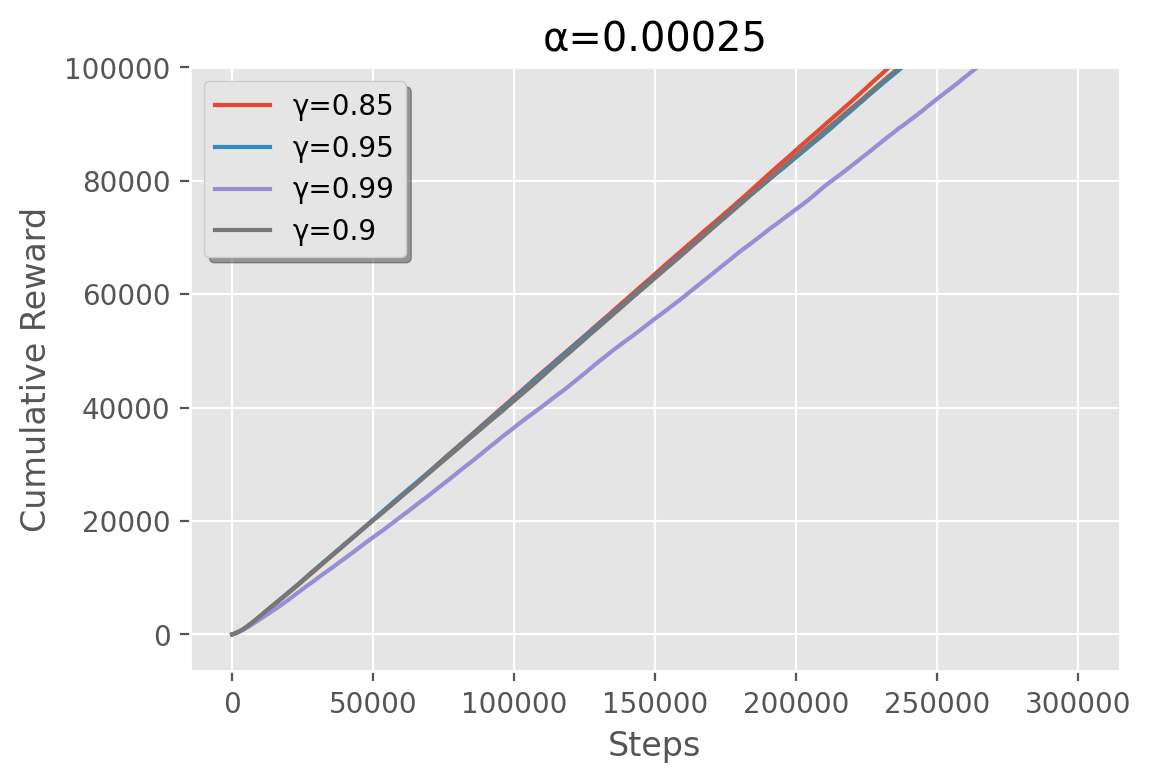}
    \caption{}
    \end{subfigure}

    \begin{subfigure}[b]{0.4\textwidth}
    \includegraphics[width=\textwidth]{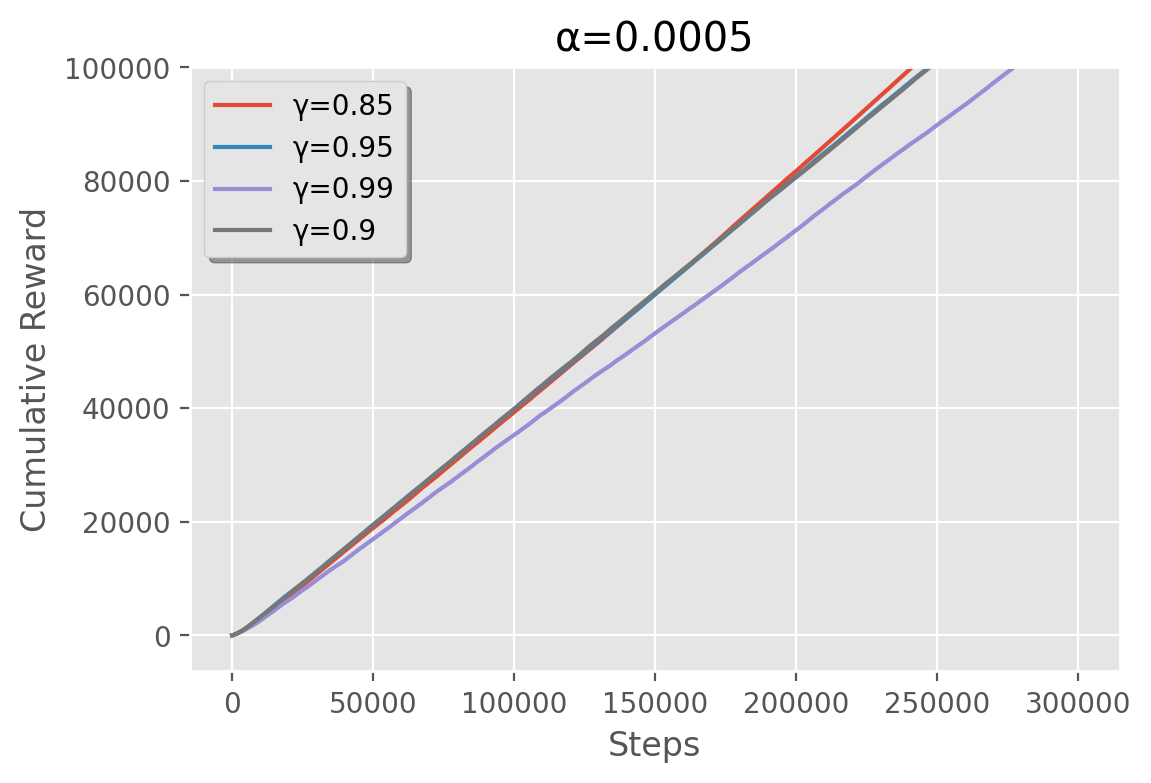}
    \caption{}
    \end{subfigure}
    \begin{subfigure}[b]{0.4\textwidth}
    \includegraphics[width=\textwidth]{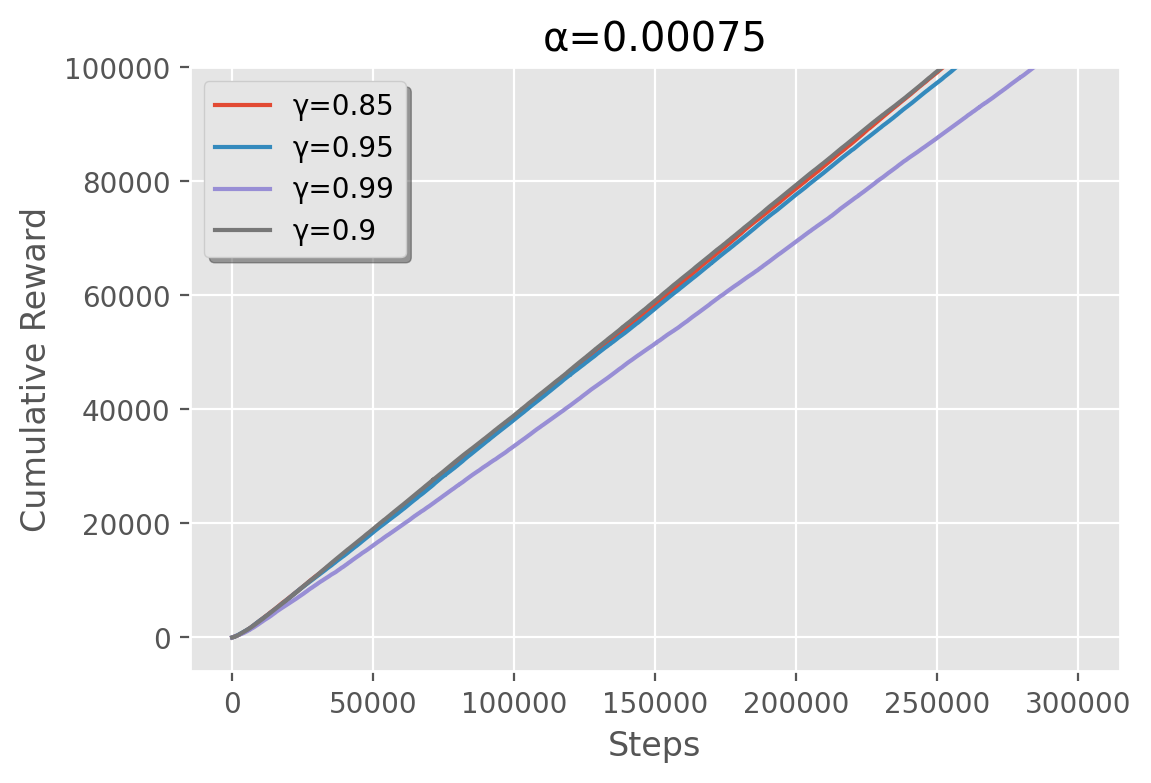}
    \caption{}
    \end{subfigure}
\caption{Parameter variation of PPO2 algorithm}
\label{fig:parametrization-ppo2}
\end{figure*}

We assume the SFC placement as an episodic task for training the algorithm, where one episode is approximately one year of customer requests. We carried out 30 experiments to train the agent with the simulation parameters presented in Table \ref{tab:simulation-parameters}. For the algorithm training, we assume 8760 steps (the number of hours in one year) multiplied by 10. To evaluate the training convergence of algorithms, we assume the cumulative reward, the sum of all rewards received so far, as a function of the number of steps. Figure \ref{fig:parametrization-ppo2} and \ref{fig:parametrization-a2c} show the parametrization results for the cumulative reward metric for A2C and PPO2, respectively.

For all combinations of parameters, both algorithms converge as we have a fast cumulative reward growth following some steps. For the A2C algorithm, the $\alpha$ value that obtained the best results was 0.00005 (Figure \ref{fig:parametrization-a2c}a), while the highest $\alpha$ value tested resulted in the worst accumulated reward results (Figure \ref{fig:parametrization-a2c}d). The discount factor $\gamma$ did not present significant impact in the cumulative reward for the A2C algorithm. All the values evaluated had very similar behaviour, but $\gamma=0.99$ resulted in better cumulative rewards. Similar to the A2C algorithm, the $\alpha$ value of 0.00005 for the PPO2 obtained the highest cumulative rewards. As shown in Figure \ref{fig:parametrization-ppo2}, a higher value of discount factor, $\gamma=0.99$, resulted in a poor convergence performance for PPO2 for all $\alpha$ values. In contrast to the A2C algorithm, $\gamma=0.85$ resulted in the greatest cumulative reward for the PPO2 algorithm.

For both algorithms, the most appropriate learning rate was 0.00005. Finding an appropriate learning rate value is an important step in adjusting the RL algorithms. As per Equation \ref{eq:q-update}, the learning rate $\alpha$ controls the adjustment made to the parameters of the policy function. Very large values can cause an exploding gradient, which compromises the agent learning. As we can see in Figures \ref{fig:parametrization-a2c} and \ref{fig:parametrization-ppo2}, the highest $\alpha$ value tested resulted in a poor convergence performance. On the other hand, the lowest $\alpha$ value resulted performed well with faster agent convergence. 

Comparing the discount factor $\gamma$, the PPO2 algorithm presented best results with a small $\gamma$, while the A2C algorithm presented the best results with higher values of $\gamma$. The discount factor determines the importance of future rewards for the current state at a given stage, which can influence the agent learning, as described in the Subsection \ref{subsec:reinforcement-learning-bkg}. Therefore, for the SFC placement problem, the A2C algorithm performed better giving less weight to future rewards, while the PPO2 algorithm worked better giving more weight to future rewards. 

When one compares the convergence velocity of both algorithms i.e. the speed at which the accumulated reward grows, the PPO2 algorithm outperformed the A2C algorithm. Figure \ref{fig:parametrization-a2c}a and Figure \ref{fig:parametrization-ppo2}a show that the cumulative reward of the PPO2 algorithm starts to grow with fewer steps than the A2C algorithm. While both algorithms achieve the cumulative reward of 100000, the PPO2 algorithm achieves this faster. After defining more suitable parameters for both algorithms, we can compare them in different SFC placement problem scenarios.
\subsection{Scenario Variation Results}
\label{subsec:secenario-variation-results}

In this subsection, we evaluate the performance of PPO2 and A2C algorithms in different scenarios, comparing them against a greedy solution (hereinafter referred to as the greedy algorithm). The greedy algorithm allocates each VNF of an SFC request following a sequential order. During the allocation, the greedy algorithm searches the server with more available resources to deploy the VNF. When the server is selected, the algorithm calculates the number of redundant VNFs to be allocated by following an uniform distribution, varying the number of VNFs from one to the maximum number of redundant VNFs allowed (see Table \ref{tab:simulation-parameters}).

For these experiments, we use a variation of the simulation setup described in Table \ref{tab:simulation-parameters}, which is showed in Table \ref{tab:simulation-parameters-variation}. We assume two groups of servers: one group where 14 servers have an MTTF of 8760 hours \cite{araujo2014availability}, and another group with 14 servers and an MTTF of 7884 hours (a reduction of 10\%). We use this variation in order to increase the server heterogeneity of the network. We also increase the customer volume to 10 in order to increase the number of SFC requests in the system. We also assume that the  availability requirement for the allocated SFCs is 99.955\%. Finally, we evaluate the scenarios for a period of five years (43800 hours). The other simulation parameters are the same as Table \ref{tab:simulation-parameters}. 

\begin{table}[h]
\caption{Variation of simulation parameters}
\centering
\label{tab:simulation-parameters-variation}
\begin{tabular}{@{}cc@{}}
\toprule
\textbf{Parameter}                 & \textbf{Value} \\ \midrule
Number of servers (Group 1)        & 14             \\
Number of servers (Group 2)        & 14             \\
Number of resources per server      &10                \\ 
Server MTTF (Group 1)              & 8760h          \\
Server MTTF (Group 2)              & 7884h          \\
Server MTTR                         &1.667h \cite{araujo2014availability} \\
VNF MTTF                            &2880h  \cite{araujo2014availability}  \\ 
VNF MTTR                            &0.17h \cite{araujo2014availability} \\ 
Energy consumption (CPU)            &40W \cite{ali2015energy}               \\
Energy consumption (memory)         &30.17W  \cite{ali2015energy}              \\
Maximum number of redundant VNFs     &4                \\
Number of Customers                & 10             \\
Availability requirement & 99.955\%       \\ 
Arrival rate ($\lambda$)            & 0.04 request/h \cite{palhares2014joint}               \\
SFC lifetime ($\mu$)                 &1000h \cite{palhares2014joint}  \\
\bottomrule
\end{tabular}
\end{table}

To evaluate the performance of the RL algorithms in the different scenarios, we train the agent for several steps, and then test it in each scenario. The number of steps to train the agent is 500000 (empirically defined). After training the RL agent, we carry out 30 simulations for each scenario configuration to evaluate the agent performance. 

\begin{figure}[h]
\centering
\includegraphics[width=0.9\columnwidth]{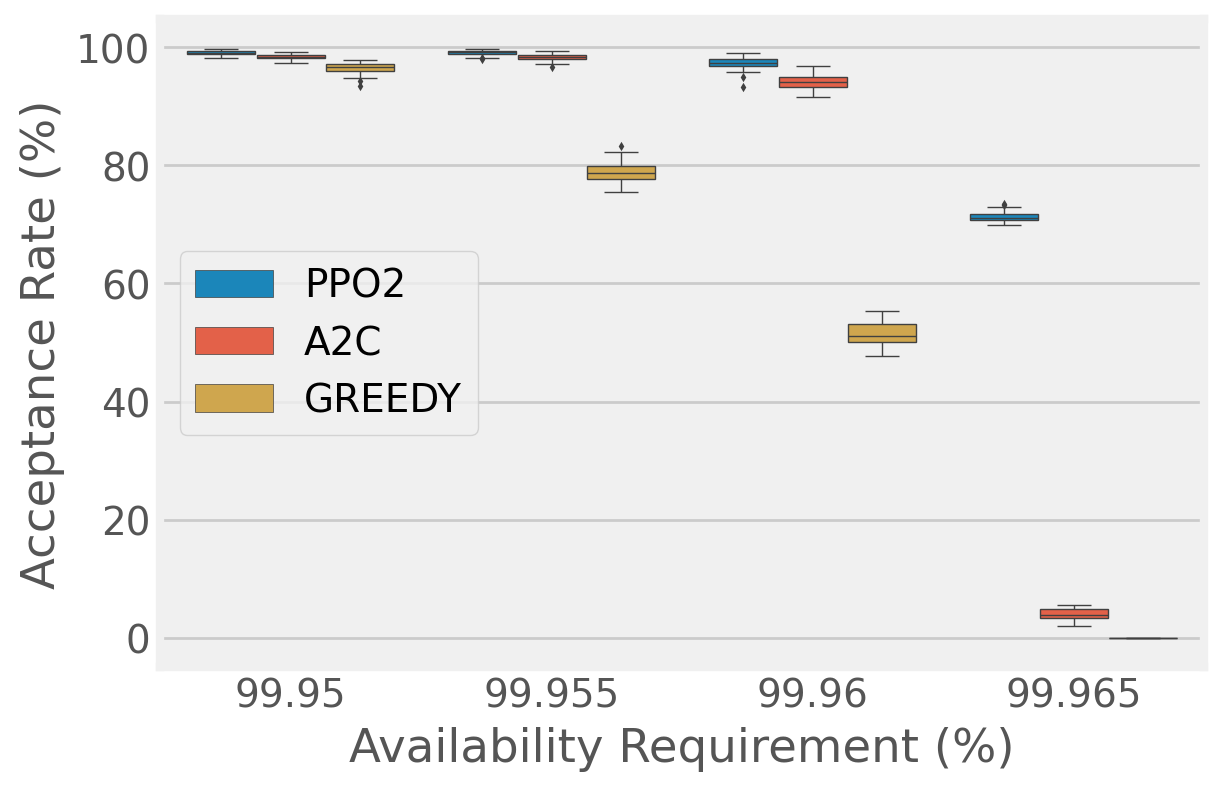}
\caption{Acceptance rate for different availability requirement levels}
\label{fig:acceptance-ratio-av-threshold}
\end{figure}

Figure \ref{fig:acceptance-ratio-av-threshold} shows the acceptance rate of algorithms for different availability requirement levels. Comparing the algorithms, the RL algorithms outperform the greedy algorithm at all availability requirement levels. When the availability requirement is 99.95\%, all techniques achieved an acceptance rate close to 100\%, with the greedy algorithm performing slightly below this level. At 99.955\%, PPO2 still achieved an acceptance rate close to 100\% while A2C performance deteriorated slightly and the  the greedy algorithm underperforming significantly at about 79\%. This suggests that the greedy algorithm is not able to adopt suitable VNF strategies to satisfy high availability requirements. At 99.96\%, the acceptance rate deteriorated  for all algorithms. The median acceptance rate was  97.33\% for PPO2, 94.10\% for A2C, and 50.01\% for the greedy algorithm. At 99.966\%, the acceptance rates for all algorithms were low. PPO2 achieved a median acceptance rate of 71.14\% while the A2C only achieved a media acceptance rate of 3.80\%; the greedy algorithm failed to allocate any SFCs at the required availability level. 

In general, as the availability requirement increases, the acceptance rate decreases considerably for all algorithms. This is because achieving a higher level of availability requires higher levels of VNF redundancy, which in turn consumes more resources from servers and makes the placement process more complex. 

The PPO2 algorithm obtained the best acceptance rate results for all availability requirement levels. This suggests that PPO2 is capable of adopting appropriate availability strategies to meet high availability requirements. On the other hand, the greedy algorithm achieved the worst results for all availability requirement values. Although the algorithm always selects the server with more available resources, the availability strategy is not optimal for meeting availability requirements. The inability of the greedy algorithm to meet high availability is particularly stark at the 99.965\% level, when no SFC was allocated. 

Comparing the two RL algorithms at the lower availability requirement levels (99.95\% and 99.955\%), both algorithms performed well with acceptance rates of approximately 98\% although PPO2 outperformed A2C. This is particularly notable at 99.96\% level, where PPO2 achieved a median acceptance rate of 97.33\% compared to 94.01\% for A2C, a difference of 3.32\%. At 99.965\%, the difference is more evident where PPO2 achieved a median acceptance rate of 71.14\% compared to 3.80\% for A2C,  a considerable difference of 67.34\%. PP02 outperforms both A2C and the greedy algorithm where customers have relatively high availability requirements i.e. 99.965\%).

\begin{figure}[h]
\centering
\includegraphics[width=0.9\columnwidth]{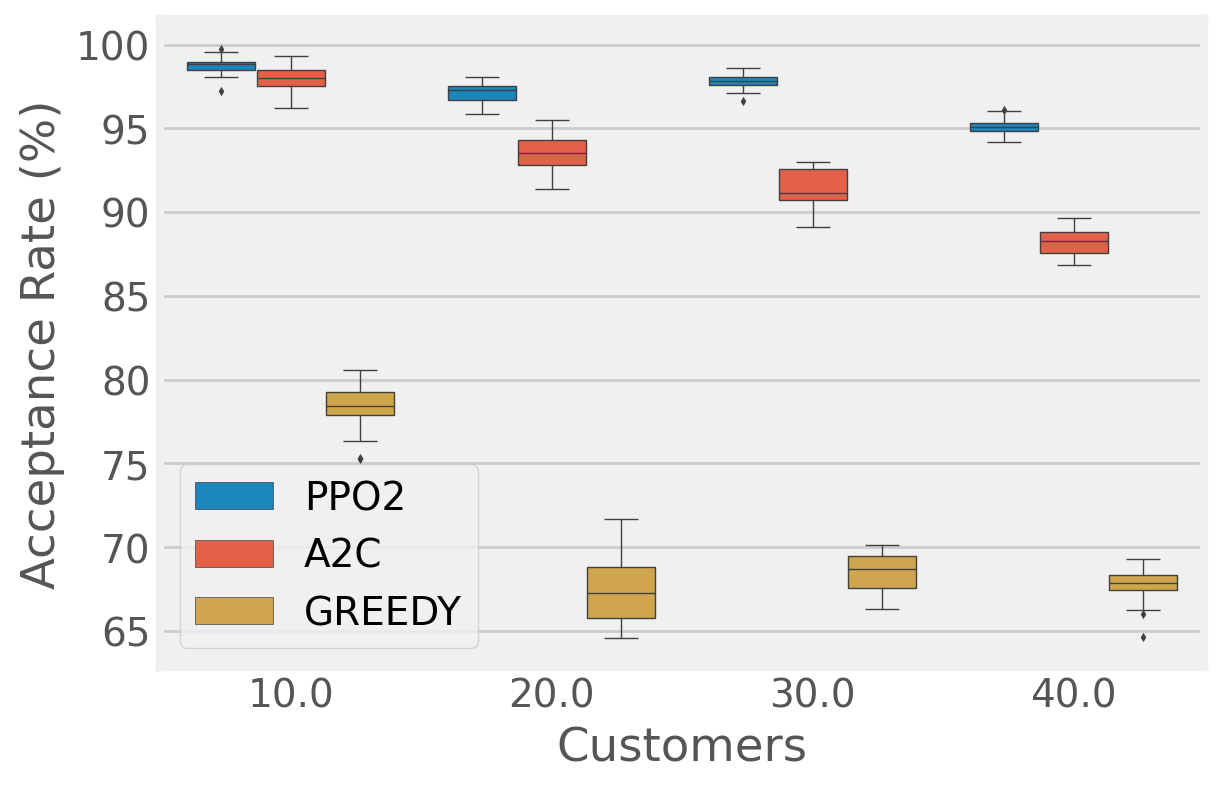}
\caption{Acceptance rates for different customer volumes}
\label{fig:acceptance-ratio-customers}
\end{figure}

Figure \ref{fig:acceptance-ratio-customers} presents the acceptance rates for all algorithms as the number of customers sending SFC requests increases. 

The greedy algorithm achieved the worst acceptance levels for all customer volumes. At 10 customers, the greedy algorithm achieved an acceptance rate of c. 79\% however when the number of customers increased, the acceptance rate deteriorated to c. 66\%. 

Comparing the RL algorithm with 10 customers, PPO2 achieved a median acceptance rate of 98.82\%, while the A2C achieved a median acceptance rate of 97.99\%, a relatively small difference of 0.83\%. At higher customer volumes, e.g. 20 and 30 customers, the is some minor deterioration in acceptance rates, c. 97\% for PPO2 and 92\% for A2C, with PPO2 outperforming the A2C in both cases. At 40 customers, the deterioration in acceptance rate is more distinct again; PPO2 achieved a median acceptance rate of 95.05\%, while the A2C only achieved 88.26\%, a difference of 6.79\%.

The performance superiority of PPO2 over both  A2C and the greedy algorithm for different customer volumes is clear. High customer volumes sending requests makes the problem hard to be solved because more SFCs must be placed, reducing the number of available resources quickly. However, when the number of customers increased from 10 to 40, the difference in the median acceptance rate decreased from 98.82\% to 95.05\% for PPO2, a difference of only 3.77\%. In contrast, both A2C and the greedy algorithm experienced a much higher deterioration in performance customer volumes and associated requests increased. Consequently, we conclude that PPO2 is the most appropriate RL algorithm of those tested for allocating SFCs as in scaling systems. 

\begin{figure}[h]
\centering
\includegraphics[width=0.9\columnwidth]{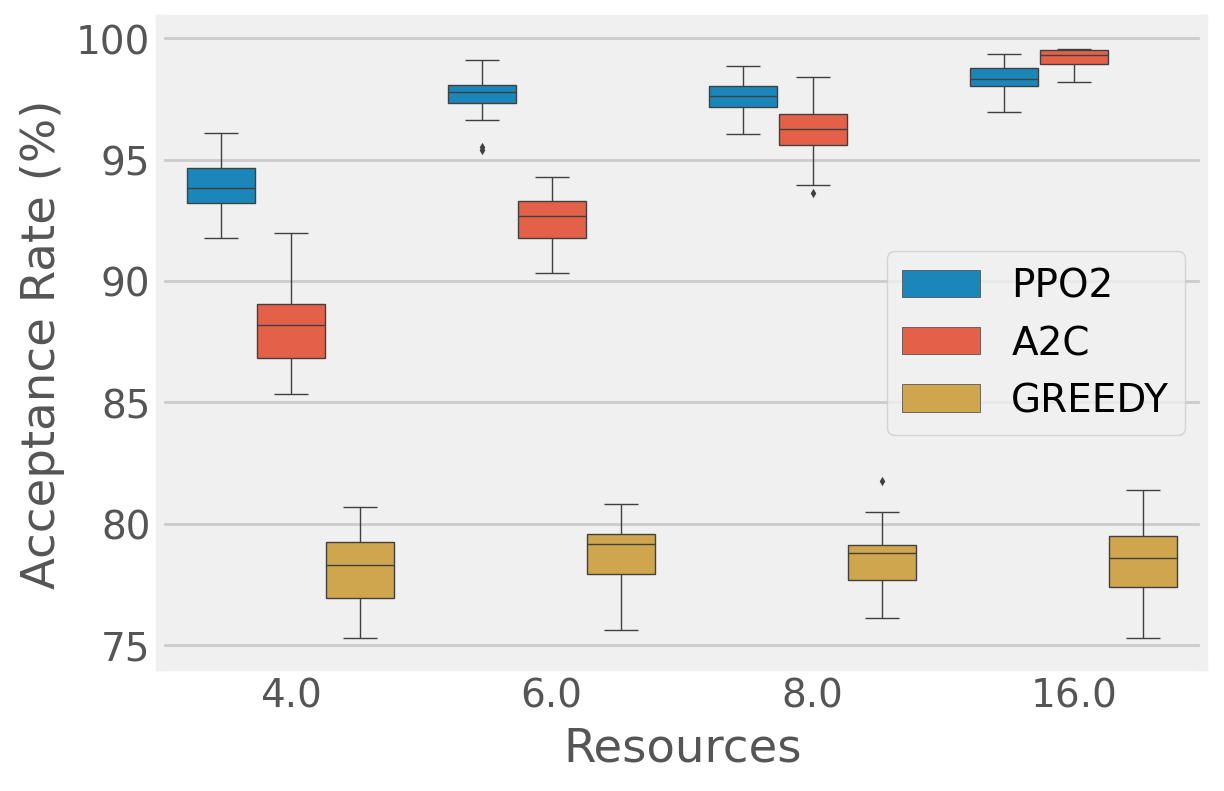}
\caption{Acceptance rate for different number of resources in the servers}
\label{fig:acceptance-ratio-resources}
\end{figure}

Figure \ref{fig:acceptance-ratio-resources} presents the acceptance rates of each algorithm for scenarios where the servers have different number of resources.

The greedy algorithm achieved a similar performance, an acceptance rate of c. 79\%, at each server resource level.
This occurs because the algorithm always finds the server with available resources. Therefore, once there are servers with resources available on the network, the algorithm will be able to allocate the SFCs. Considering the number of SFC requests generated in the simulation, the number of server resources was not a limiting factor for the allocation. As discussed earlier, the limiting factor is the availability requirements as set our in the SLA and agreed with the customer. The greedy algorithm is not able to meet these requirements even with more resources in the servers.

Regarding the performance of the RL algorithms, where the servers had between four and eight resources, the PPO2 outperformed the A2C algorithm. However, as the number of resources available in the server increased, the performance of PPO2 and A2C differ. At 16 resources per server, the A2C surpassed PPO2, achieving a better acceptance rate. For instance, assuming four resources in the servers, the median acceptance rate was 93.85\% for the PPO2 and 88.20\% for the A2C, a difference of 5.65\%. For eight available resources in the servers, the median acceptance rate for PPO2 increased to 97.61\% and to 96.24\% for the A2C, a difference of 1.37\%. For the scenario where the servers have 16 resources, the PPO2 achieved a median acceptance rate of 98.32\%, while the A2C achieved 99.32\%. While a minor difference, it indicates a positive trend for A2C. As the number of resources in the servers increased, the acceptance ratio of RL increased as expected, since more resources were available to allocate the SFCs. However, as presented in Figure \ref{fig:acceptance-ratio-resources}, as the number of server resources increased, the acceptance rate of PPO2 grew from around 94\% to around 98\% and stabilised. However, the acceptance rate of A2C had a linear growth as the number of resources increased. Therefore, one can conclude that for environments where  servers have more computational resources, A2C is the more appropriate, while for scenarios where servers have less resources, the PPO2 is the algorithm with highest acceptance rate.

We also evaluated the energy consumption for each allocated SFC based on the assumptions in Table \ref{tab:simulation-parameters-variation}. Table \ref{tab:energy-consuption-results} presents the results. PPO2 presented the lowest mean energy consumption at 136.81W, followed by A2C at 137.82W, and the greedy algorithm at 140.34W. However, based on the standard deviation, the energy consumption of PPO2 and A2C can be considered statistically similar. Therefore, in addition to higher acceptance rate values, the RL algorithms are also able to achieve greater energy efficiency in allocating SFCs than the greedy algorithm. If only the average value is assumed, PPO2 is able to deliver greater energy efficiency than the others in addition to a higher acceptance rate in most of the evaluated scenarios.

\begin{table}[h]
\centering
\caption{Energy consumption per allocated SFC}
\label{tab:energy-consuption-results}
\begin{tabular}{@{}ccc@{}}
\toprule
\textbf{Algorithm} & \textbf{Mean Energy Consumption (W)} & \textbf{Standard Deviation} \\ \midrule
PPO2               & 136.81                                  & 0.74                        \\
A2C                & 137.82                                  & 0.72                        \\
Greedy             & 140.34                                  & 1.61 $\times 10^{-13}$                    \\ \bottomrule
\end{tabular}
\end{table}

\section{Related Works}
\label{sec:related-works}
There is a growing literature on all aspects of SFCs \cite{bhamare2016survey,de2019network}. This article straddles three literature bases related to SfC placement - RL,  availability, and energy consumption. While each domain in themselves has established work, few studies address the use of RL for availability-aware and energy-aware SFC placement.

\subsection{Reinforcement learning-based SFC Placement}
The authors in \cite{xiao2019nfvdeep} proposed a system (NFVdeep) based on deep RL (DRL) to minimize the operation cost of NFV service providers and maximize the total throughput of requests, taking into account several QoS requirements. They used a policy gradient approach to solve the large discrete action space problem. The results suggest that NFVdeep reduced costs by 33.29\% and increased accepted throughput by 32.59\%.
The authors in \cite{chai2019parallel} proposed a parallel SFC placement (PP-DRL) scheme based on
DRL and a heuristic algorithm to allocate resources
proportionally.  The main difference is that different SFC requests can share the VFNs created to reduce resource consumption and resource cost. By elastically adapting to a wide variety of networks, it was hypothesised that PP-DRL better utilises fewer resources to serve a larger size of demands. Results suggest that the proposed solution outperformed reference studies with respect to acceptance ratio, and both link resource cost and VM resource cost. The work presented in \cite{luo2019scaling} proposes a DRL-based framework that combines a recurrent neural network (RNN) with a RL agent to cost-efficiently deploy SFCs onto geographically distributed data centers. The RNN takes as input a historical time-window of traffic flow and predicts the flow rate in the next time slot. Then, based on the current placement of the VNF chain and the predicted flow rate, produces VNF placement decisions for the deployed SFC. Results suggest that the solution proposed outperforms other schements with a cost reduction of up to 42\%. In \cite{guo2019trusted}, the authors seek to address trust and adaptation issues in resource allocation in heterogeneous architectures found in cloud-edge networks. The authors propose an SFC orchestration architecture based on consortium blockchain and DRL.  The asynchronous advantage actor-critic (A3C) algorithm is used for SFC placement. A3C has a practical advantage in that it can run on a multicore CPU independently which presents faster execution times compared to alternatives\cite{guo2019trusted}. In simulations, the A3C-based solution outperformed DQN and the Link Stating Routing Algorithm (LRSA) both in terms of delay, overall network cost. 
In \cite{khezri2019deep}, a DRL-based solution was proposed for SFC placement considering the reliability requirements from customers. Reliability is the probability of a system operates without failures until a certain time\cite{kumar2017system}. The authors propose an agent based on DQN for placement to maximise the acceptance ratio while, at the same time, minimising placement costs and meeting reliability constraints set by customers. 

Although, the studies above are relevant to SFC allocation and consider important metrics such as SFC cost, delay, and throughput; none of these works consider the availability of allocated SFCs. In this paper, we consider the availability of both hardware servers and VNFs in the SFC placement. In addition, we consider the energy consumption of the VNFs in order to avoid redundant NFVs that consume energy unnecessarily.

\subsection{Availability-aware SFC Placement}
A number of works examine availability aspects in SFC placement using strategies other than RL.The authors in \cite{fan2017availability} proposed an algorithm to deploy redundant VFNs to increase SFC availability. The algorithm takes into account the available resources, the link capacity, the overall delay, and the customer availability requirements. Then, for each SFC request, a greedy algorithm with a theoretical lower bound selects the minimum number of redundant VFNs needed, deploys the VFNs in the data center, and maps the logical links to meet all constraints. Results suggest that the proposed algorithm reduced the resource consumption and increased the request acceptance ratio when compared to other papers from the literature. In \cite{moualla2018availability}, the authors proposed an Integer Linear Programming (ILP) solution for SFC placement for data centers with Fat-Tree topologies. The authors proposed an algorithm for the placement that considers the availability of servers and switches present in the data center, customer availability requirements, and server utilisation. Results suggest that the algorithm is able to satisfy the customer availability requirements while also improving the network server CPU utilisation. In \cite{he2019asco}, the authors proposed Availability-ware SFC Orchestration (ASCO) for SFC placement. ASCO is based on a pre-pruned depth-first search (PDFS) algorithm. The algorithm is designed to cost-efficiently meet QoS and availability requirements for SFC placement. The results suggested that ASCO outperformed other algorithms (based on the minimum cost and single path selection) from the literature based on acceptation rate, total cost, node usage, and link usage. 

While the availability-aware studies presented above have similarities to our study,  there are some important differences. The works presented in \cite{moualla2018availability} and \cite{kumar2017system} considered a static infrastructure, where components do not fail. In a real scenario, server and software failures happen thereby compromising SFC placement. By considering failure and repair rates, we consider a more dynamic environment with a higher level of verisimilitude. In our study, the agent considers both server suitability and the redundancy strategy for each VNF in an SFC. In addition, we considered availability and energy consumption for the SFC placement. The availability is estimated through RBD models and the energy consumption is calculated based on the servers where the SFC is placed. These SFC aspects are considered for the RL agents during their training in order to learn a SFC placement strategy which met the availability requirements from customers and reduce the SFC energy consumption.

\subsection{Energy-aware SFC placement}
\label{subsec:energy-aware-sfc-placement}
A number of studies explore energy consumption optimisation for various SFC placement scenarios. Kouah et al. \cite{kouah2018energy} formulated the SFC placement problem as a Mixed Integer Program with the goal of minimising energy consumption for the Internet of Things (IoT), where form factors can result in significant energy limitations. a use case which scenarios. Results suggest that proposed solution provided an optimal solution for sensing, processing and communication energy consumption in small network topologies and extending the lifetime of target devices. Sun et al. \cite{sun2019energy} explored energy consumption optimisation of SFC placement in a multi-cloud scenario. The authors formulated the SFC placement problem as an ILP and use a low-complexity heuristic algorithm for SFC orchestration across multiple domains to near-optimally solve this problem. The proposed solution outperforms existing approaches, namely the Nestor algorithm and the DistNSE algorithm, for both power consumption and average response time.
The works presented in \cite{xu2018energy} and \cite{zhang2019near} focused on the SFC placement in telecommunication networks. In both instances, energy consumption optimisation was formulated as an ILP taking into account the links and server capacities. A Markov-based algorithm was proposed to find the near-optimal solution in a polynomial time. In both papers, performance was compared with a greedy algorithm and simulation results suggest that the proposed Markov-based algorithm was significantly more energy efficient within acceptable periods of time.

Although these works address energy consumption during the SFC placement, this paper differs significantly in two ways. Firstly, while the previous works considered heuristics in their solutions, we assume the SFC energy consumption as a metric for a RL agent, who learns to minimise such consumption after the learning process. Secondly, we consider two potentially conflicting factors - availability and energy efficiency - during SFC placement.  Thus, after the learning process, the RL agent is able to allocate the SFC based on the availability levels required by the SLA and/or the customer while at the same time minimising energy consumption to arrive at an optimal solution. 

\section{Conclusion and Future Works}
\label{sec:conclusion}
In this paper, we proposed the use of RL agents to place SFCs taking into account availability requirements and energy consumption constraints. We formulated the SFC placement problem as an MDP process, where the SFC requests are processed sequentially to be deployed in commodity servers. These servers may fail during their operation exacerbating the complexity of the placement problem. We proposed the use of RL for the SFC placement. The SFC request is composed of the VNFs required by the customer and an availability level requirement which must be met by the service provider. Each VNF has also computing resources requirements.


We developed a simulator to compare two policy-based RL algorithms, the PPO2 and A2C. We carried out experiments to find the best parameters for both algorithms. We then used the best parameters to evaluate the algorithms in different scenarios, comparing against a greedy approach. The results showed that the RL algorithms outperformed the greedy approach for all scenarios evaluated, both in terms of acceptance rate and energy consumption. The experiments suggested that PPO2 presented better performance than A2C in most of the scenarios. Therefore, we can conclude that the PPO2 is the more appropriate technique for SFC placement taking into account SFC availability and energy consumption. However, we note that where more resources per server are available, A2C presents promising results.

For future research, we intend to explore other relevant elements of SFC placement and add delay and link bandwidth, as well as QoS requirements. Since RL algorithms can process the SFC requests in an online manner, we plan to use the traffic flow of SFCs in the placement. This will enable a more granular consideration of server suitability for the respective VFNs. Finally, we also plan to evaluate other RL techniques, identifying their optimal parameter configuration for different SLA requirements and scenarios.

\bibliographystyle{unsrt}  
\bibliography{references}  






\end{document}